# High-level synthesis design of scalable ultrafast ultrasound beamformer with single FPGA

Zhengchang Kou, Qi You, Jihun Kim, Zhijie Dong, Matthew R. Lowerison,
Nathiya V. Chandra Sekaran, Daniel A. Llano,
Pengfei Song, Senior *Member, IEEE*, Michael L. Oelze, Senior *Member, IEEE*

*Abstract*—Ultrafast ultrasound imaging is essential for advanced ultrasound imaging techniques such as ultrasound localization microscopy (ULM) and functional ultrasound (fUS). Current ultrafast ultrasound imaging is challenged by the ultrahigh data bandwidth associated with the radio frequency (RF) signal, and by the latency of the computationally expensive beamforming process. As such, continuous ultrafast data acquisition and beamforming remain elusive with existing software beamformers based on CPUs or GPUs. To address these challenges, the proposed work introduces a novel method of implementing an ultrafast ultrasound beamformer specifically for ultrafast plane wave imaging (PWI) on a field programmable gate array (FPGA) by using high-level synthesis. A parallelized implementation of the beamformer on a single FPGA was proposed by 1) utilizing a delay compression technique to reduce the delay profile size, which enables both run-time pre-calculated delay profile loading from external memory and delay reuse 2) vectorizing channel data fetching which is enabled by delay reuse, and 3) using fixed summing networks to reduce consumption of logic resources. Our proposed method presents two unique advantages over current FPGA beamformers: 1) high scalability that allows fast adaptation to different FPGA resources and beamforming speed demands by using Xilinx High-Level Synthesis as the development tool, and 2) a compact form factor design by using a single FPGA to complete the beamforming instead of multiple FPGAs. Current Xilinx FPGA provides the capabilities of connecting up to 1024 ultrasound channels with a single FPGA and the newest JESD204B interface analog front end (AFE). This channel count is much more than the channel count needed by current linear arrays, which normally have 128 or 256 channels. With the proposed method, a sustainable average beamforming rate of 4.83 G samples/second in terms of input raw RF sample was achieved. The resulting image quality of the proposed beamformer was compared with the software beamformer on the Verasonics Vantage system for both phantom imaging and *in vivo* imaging of a mouse brain. Multiple imaging schemes including B-mode, power Doppler and ULM were assessed to verify that the image quality was not compromised for speed.

## I. INTRODUCTION

Ultrafast ultrasound [1],[2] has been the driving force for many new ultrasound imaging applications such as shear wave elastography[3],[4], super-resolution ultrasound localization microscopy (ULM) [5],[6], and functional ultrasound (fUS) [7]. Traditional line by line or multi-line focused beam scanning cannot provide sufficient frame rates for applications requiring high-speed tracking of tissue motion (e.g., blood flow, shear wave motion, microbubble movement) within a large field-of-view (FOV). Compound plane wave imaging (CPWI) [8] provides a balance between imaging quality (e.g., spatial resolution, signal to noise ratio (SNR)) and imaging frame rate and is widely used in applications that require ultrafast ultrasound.

Ultrafast ultrasound with CPWI is a demanding technology with high data rate and beamforming computational cost. Ultrafast ultrasound typically requires a peak data rate at 112 Gb/s and a peak sample rate at 8 GSPS for a 128-channel system with a 14-bit, 62.5 MHz analog to digital converter (ADC). To achieve continuous non-blocking ultrafast ultrasound imaging, the beamformer's sustainable average beamforming rate is not necessary to meet the peak data rate because the duty cycle (acquisition time/acquisition interval) of the ultrasound data acquisition is not 100%. Depending on the application, this duty cycle normally ranges from 10% to 30% [6],[7],[9]. State-of-the-art ultrafast imagers such as the Verasonics system use CPU-based software beamformers with PCI-Express interface to handle raw radiofrequency (RF) data transfer [10]. However, the sustainable average beamforming rate of Verasonics' CPU beamformer is limited to 133 M samples/second (MSPS), according to our measurements. For applications such as fUS and ULM, where long data acquisitions with high frame rate are necessary, CPU-based beamforming does not provide adequate sustainable average beamforming speed. As a result, users have to choose between recording unbeamformed RF data to local storage (for offline beamforming) with small time gaps between data blocks (i.e., continuous data acquisition within each block) [11], or recording beamformed data with a relatively large time gap between consecutive time blocks [12]. As compared to a CPU, GPU-based beamformers [13] provide significantly improved beamforming speed [14]. However, due to the data transfer overhead, when using a GPU, more than half of the

This work was supported in part by the by the National Institutes of Health under grants R21EB024133, R21EB023403, R21EB030743, R01CA251939, R21EB030072, R01EB031040, R21AG077173, and in part by the National Science Foundation Award 2237166 (Corresponding author: Michael L. Oelze.)

All authors are affiliated with the Beckman Institute for Advanced Science and Technology, University of Illinois at Urbana-Champaign, Urbana, IL 61820 USA (email: zkou2@illinois.edu; oelze@illinois.edu); In addition, Z. Kou, J. Kim, Z. Dong, P. Song and M.L. Oelze are with the

Department of Electrical and Computer Engineering; P. Song, D. Llano, and M. Oelze are with the Carle Illinois College of Medicine; Q. You, P. Song, and M. L. Oelze are with the Department of Bioengineering; Nathiya V. Chandra Sekaran and Daniel A. Llano are with the Department of Molecular and Integrative Physiology, University of Illinois Urbana-Champaign, Urbana, IL 61801 USA J. Kim is with the Division of ICT Convergence Engineering/Major in Electronic Engineering, Kangnam University, Republic of Korea.

beamforming time is consumed in data transfer from the Verasonics to the host PC. As such, the bandwidth of the PCI-Express interface remains the key barrier for achieving continuous data acquisition at an ultrafast imaging frame rate [15]. Recently, the GPU Direct option released by Verasonics reduced the transfer time allowing 2 GB RF data to be transferred directly to the GPU RAM in a little over 300 ms.

To overcome the existing challenges associated with software beamformers, many efforts have been made to implement the beamformer on field programmable gate array (FPGA). An FPGA-based beamformer could provide unique potential advantages: 1) an FPGA can directly interface with the analog front end (AFE) chips, which makes the data transfer overhead negligible between the ultrasound data acquisition module and the beamformer, and 2) an FPGA supports massive parallel beamforming with much higher computational performance and lower power consumption thanks to its fully programmable memory and computational architecture. Compared to a CPU, a much higher sustainable average beamforming rate could be achieved by a FPGA. Therefore, the FPGA provides an enticing solution for continuous recording and beamforming of ultrafast ultrasound data with a speed that is only limited by the speed of sound propagation in soft tissue.

Some existing ultrasound systems, such as the ULA-OP 256, are FPGA-based [16],[17]. In their design, FPGAs are directly connected to AFEs for beamforming. Groups of 32 elements are connected to one FPGA beamformer, which has four DAS units inside. Each DAS unit individually beamforms one line of an image. A continuous beamformed pulse repetition frequency (PRF) of 3,800 Hz was achieved with this system. The beamformer could produce up to 467 MSPS which is much higher than 133 MSPS achieved by the Verasonics CPU beamformer. Even though the frame rates achieved by FPGA-based beamformers [17] are much higher than for CPU and GPU beamformers, the achievable continuous frame rate is still limited and cannot match with the front-end acquisition frame rate (e.g., 10,000 Hz). A more recent study utilized 8 Xilinx Kintex 7 FPGAs for parallel beamforming which achieved a beamforming rate up to 917 MSPS for a 64-channel system [18].

To improve the FPGA beamformer performance and implementation efficiency, several different methods were proposed to ease the pressure on the delay calculation, which is expensive to implement on a FPGA as it includes the square root operation. Directly loading the pre-calculated delay profile to a FPGA at run-time is not feasible due to the enormous size of the delay profile [19]. One solution is to reduce the calculation cost by approximation to enable run-time delay calculation on FPGA. An efficient approximated algorithm [20], which is based on the first order Taylor expansion of the square root, was proposed to reduce the delay calculation cost on a FPGA. A portable 3D beamformer with divergent wave transmission [21] was built based on this method. The other solution is to compress the delay profile to enable the pre-calculated delay profile loading at the run-time [18]. A delay compression technique that recognizes additional symmetry between the round-trip times for all focal points drastically reduces the delay profile size [22] and a low power 3D beamformer has been designed on an application-specific integrated circuit (ASIC) based on this method [19].

One of the advantages of FPGA over ASIC is it allows users to reprogram the device to fit different applications. While the traditional hardware description language (HDL) usually requires a longer development time compared to the software programming, high-level synthesis (HLS) could effectively reduce the length of the time to market by allowing faster algorithm verification in a FPGA compared to the traditional HDL. A previous study demonstrated that HLS could be used to implement a FPGA beamformer for synthetic aperture (SA) imaging [23].

In this paper, we propose an HLS designed novel FPGA-based hardware beamformer with optimized hardware parallelization implementation which is enabled by an efficient delay profile compression algorithm. In our design, 4.83 G samples/second (GSPS) beamforming rate is achieved in terms of input raw RF samples. The delay compression enables delay reuse and run-time precalculated delay profile loading from external DDR memory. As the same delay profile is reused for all the pixels at the same depth, the proposed beamformer could beamform one row of an image simultaneously. Further, memory access for raw RF data is in a 2,048 bits wide vector scheme to increase memory utilization efficiency. Multiple buffers with identical data are used to provide over 9 Tbps internal memory bandwidth.

## II. METHODS

### A. Delay and Sum (DAS) beamforming background

DAS and its limitation to parallelization is introduced in this section. The underlying principle of DAS is to achieve dynamic receive focusing by compensating both the transmit and receive delays for each imaging pixel. The traditional delay calculation of DAS is illustrated in Fig.1. The transmit delay of a plane wave at steering angle $\theta$ to reach a pixel point at $(x, z)$ is

$$\tau_{tx}(\theta, x, z) = \frac{z * \cos\theta + x * \sin\theta}{c}, \quad (1)$$

where $c$ is the speed of sound. The receive delay, which corresponds to the time required by the signal reflected from $(x, z)$ to reach a receive element at $x_n = (n - 1) * d$, is

$$\tau_{rx}(x_n, x, z) = \frac{\sqrt{z^2 + (x_n - x)^2}}{c}, \quad (2)$$

where $n$ is the index of the element and $d$ is the element spacing.

From Eq. 1 and 2 we can observe that the total delay is dependent on both the lateral and axial positions of the target pixel and the lateral position of the receive element. To simplify the discussion, we define the total number of elements (ultrasound channels) as $W_i$, the number of elements used to beamform an A-line as $F$ (subaperture size which is no larger than $W_i$), the total number of RF samples along the fast time (axial direction) within one frame as $D$ and the number of A-lines output from the beamformer as $W_o$. To further simplify the discussion at the beginning, we set the $W_o = W_i$, which means the number of output A-lines is the same as the number of elements.

The first limitation resides in the size of the delay profile. To beamform one frame with the traditional delay calculation, we need to calculate the delay matrix $\tau_{rx}$ with the size of $D * W * F$. If 16 bits is used to store each delay value, which is used as a raw RF data address, the delay profile for each steering angle could consume 320 Mb of space given $D$, $W$ and $F$ are 2560, 128 and 64, respectively, for one of the transducer probes used in this work. The required space for the full delay profile is too large to have the delay profile pre-calculated and loaded to the FPGA's on-chip memory before the beamforming. So, either delay compression with run-time delay calculation, loading from external DDR memory methods or run-time delay calculation methods are needed to enable FPGA based DAS beamformer.

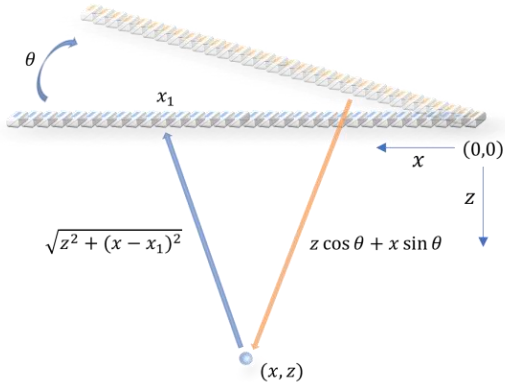

**Figure 1. Schematic plot of transmit and receive delay calculation used in DAS beamforming for a steered plane wave.**

The second limitation resides in directly reading raw RF data needed by the beamformer from external DDR memory, which could easily exceed the DDR memory bandwidth for ultrafast imaging. For example, aiming at 10,000 frames per second (FPS) beamforming rate with the parameter defined previously, the required memory bandwidth for direct reading is 3.36 Tbps (2560*128*64*10000*16 bits) which is much higher than current DDR4 memory bandwidth. Reading raw RF data from on-chip internal memory resources, such as the block RAM (BRAM), which functions as a buffer between the external DDR memory and the beamformer, becomes a better choice as thousands of BRAMs are available in current FPGAs that could provide over Tbps internal memory bandwidth.

After reading the raw RF data from the internal memory resources, all the raw RF data needed by one image pixel are summed together to finalize DAS.

### B. Delay profile compression and reuse

To improve the parallelization efficiency, we adopt the delay compression methods introduced in [22] and modify it for FPGA implementation.

We first reduce the dimensions of the receive delay profile matrix $\tau_{rx}$ from three to two by

$$\Delta x = x_n - x \qquad (3)$$

which represents the relative lateral distance between the pixel and the receive element. The maximum range of $\Delta x$ is the subaperture size $F$. Then, the revised receive delay can be rewritten as

$$\tau'_{rx}(\Delta x, z) = \frac{\sqrt{z^2 + \Delta x^2}}{c}. \qquad (4)$$

In this way, we do not calculate the delay profile according to both the lateral and axial position of the pixel and the lateral position of the receive element. Instead, we calculate the delay profile according to the lateral distance between the pixel and the receive element and the axial position of pixel.

Following the same principle, the transmit delay is revised to:

$$\tau'_{tx}(\theta, x_n, \Delta x, z) = \frac{z * \cos\theta + (x_n - \Delta x) * \sin\theta}{c}. \qquad (5)$$

To remove $x_n$ from the transmit delay calculation, we can remove the beginning $N_n^{remove}$ samples of RF data in the receive channel $n$ with the number of removed samples calculated from Eq. 5

$$N_n^{remove} = \frac{x_n * \sin\theta * f_s}{c}, \qquad (6)$$

where $f_s$ is the RF sampling frequency. Then, we can rewrite Eq. 5 as

$$\tau''_{tx}(\theta, \Delta x, z) = \frac{z * \cos\theta - \Delta x * \sin\theta}{c}. \qquad (7)$$

Equations 4 and 7 can be combined to write the total delay as

$$\tau_{total}(\theta, \Delta x, z) = \tau''_{tx}(\theta, \Delta x, z) + \tau'_{rx}(\Delta x, z). \qquad (8)$$

In Eq.7, we can observe that now the delay is only dependent on the steering angle $\theta$, relative lateral distance $\Delta x$, and depth $z$. By using Eq.7, we can reduce the delay profile for each angle from a 3D matrix ($D * W * F$) to a 2D matrix ($D * F$). Besides, given a fixed steering angle and depth, the only variable in the delay calculation is the relative lateral distance between the pixel and the receive elements. As a result, all the pixels at the same depth use the same set of delay values. In other words, we can use reuse one set of delay values ($F$ delay indexes) for one pixel to beamform all the pixels at the same depth.

### C. Parallelization

Parallelization occurs in two parts: memory parallelization and DAS parallelization.

The first step of memory parallelization is based on the delay reuse introduced in the previous section. As all the pixels at the same depth use the same set of delay indexes ($F$ delay indexes), we can use each delay index to read the input RF data samples at the corresponding depth from all the input channels (one row of RF data samples) instead of single input channels (single RF data sample). In this way, the input RF data samples are vectorized. Each delay index is used as a pointer to a vector that is $W_i$ elements wide instead of single RF data samples.

The second step of memory parallelization is to read all the delay indices ($F$ delay indexes) at the same depth, as Fig. 2(a) shows. Then, $F$ rows of input RF data samples ($F$ vectors of $W_i$

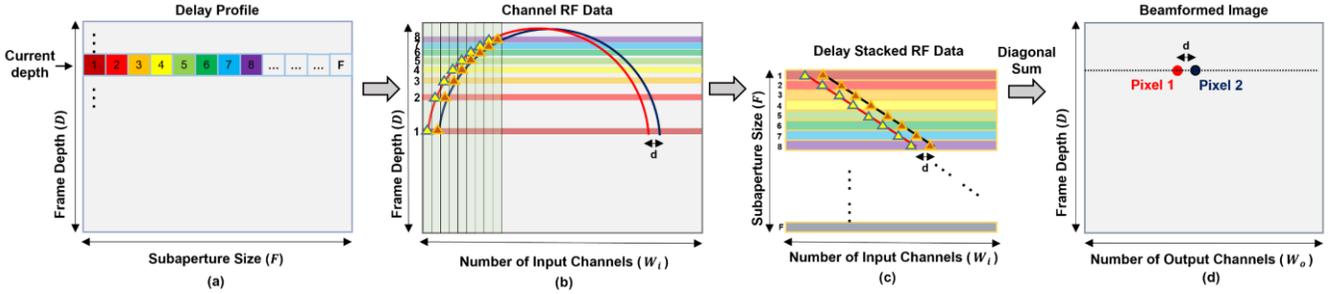

**Figure 2. Block diagram of the proposed parallelized DAS algorithm. (a)** Reading delay indexes from the delay profile generated by the revised delay calculation algorithm. Given the current depth to beamform, all the delay indexes at this depth are fetched ($F$ delay indexes). **(b)** Reading $F$ rows of RF data according to the delay indexes fetched in the previous step. The diagram only demonstrates the first 8 delay indexes and the corresponding 8 rows of RF data to simplify the diagram. **(c)** Stacking $F$ rows of RF data from the previous step together. Diagonal sums are performed on the stacked RF data to beamform each pixel at the current depth. **(d)** Two beamformed pixels are highlighted in the beamformed image.

elements wide) are read from the channel RF data buffer, as Fig. 2(b) shows.

After reading $F$ rows of input RF data samples in the memory parallelization, the DAS parallelization is performed by first stacking $F$ rows of input RF data samples together following the order in the delay profile as Fig. 2(c) shows. Then, the diagonal sum is performed to generate the beamformed result, as Fig. 2(d) shows.

The prerequisite of this DAS parallelization is that both the number of beamformed lines and the pitch ($d$) of the proposed beamformer's output are the same as those of the input ultrasound channels. As a result, $\Delta x$ also has a step size of $d$. In this way, the lateral spacing of two laterally adjacent delay indices in the delay profile, two adjacent A-lines and two adjacent input channels, are all $d$. Then, we can derive two properties. First, the RF data samples, which are needed by the beamforming of one pixel, are aligned as a diagonal line in the delay stacked RF data matrix. This is because both the $\Delta x$ and input channels ($x_n$) have a lateral spacing or step size of $d$. When both are incremented by the same step size $d$, the $x$ remains unchanged (targeting pixel position unchanged) according to Eq. 3. This simultaneous increment corresponds to the diagonal direction in the delay stacked RF data matrix in Fig. 2(c). For example, in Fig. 2(c), the red diagonal line corresponds to all the RF data samples needed by the beamforming of one pixel.

Second, two adjacent targeting pixels can be beamformed from two adjacent diagonal lines in the delay stacked matrix. This is because when both $x$ and $x_n$ are incremented by the same step size $d$, $\Delta x$ remains unchanged according to Eq. 3, which means two adjacent RF data samples in the same row in Fig. 2(c) individually contribute to two laterally adjacent beamformed pixels in Fig. 2(d). Combining this with the first property, we can conclude that all the targeting pixels at the same depth can be beamformed by diagonally summing up the delay stacked RF data matrix. For example, the red and blue pixels in Fig. 2(d) have a lateral distance of $d$. The corresponding RF data samples needed by the beamforming of these two pixels are two diagonal lines which also have a lateral spacing of $d$.

One thing to note is that fixing both the number and the pitch of the output A-lines from a proposed beamformer does not necessarily mean we have to have the number of receive channels the same as that of receive channels, because $x$ can start from $\frac{d}{2}$, $\frac{d}{3}$ or $\frac{d}{4}$ and so on. As a result, we can use multiple identical beamformers that are loaded with different delay profile basing on different initial $x$ to reconstruct images with finer spatial pixel resolution.

By performing these two parallelization operations, in the ideal case we only need to read one row ($F$ delay indexes) from the delay profile matrix and load the corresponding $F$ rows of raw RF data to the stacked RF data matrix to beamform all pixels at the same depth, with all the computation being fully parallel. While due to the FPGA's resource limitation, we need to separate this fully parallel process into multiple sequential process which is described in the next section.

*D. Implementation*

To implement the beamformer on the FPGA, the first step was to perform a linear interpolation by a factor of two on the raw RF data in the fast time direction to improve accuracy of the delay profile. To implement the linear interpolation, a local buffer was used to buffer two consecutive rows of raw RF data from all the input channels. Then, the interpolated row was generated by calculating the mean of these two rows. The output of linear interpolator was reordered to output the interpolated row between two raw RF data rows.

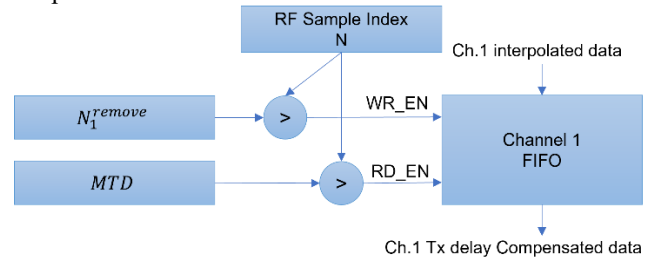

**Figure 3. Block diagram of the proposed transmit delay compensation mechanism.** This diagram shows the transmit delay compensation for input channel 1. All the other input channels have the same structure. The FIFO receives the write enable (WR_EN) signal when current RF sample index $N$ is larger than $N_1^{remove}$ and receives the read enable (RD_EN) signal when $N$ is larger than $MTD$.

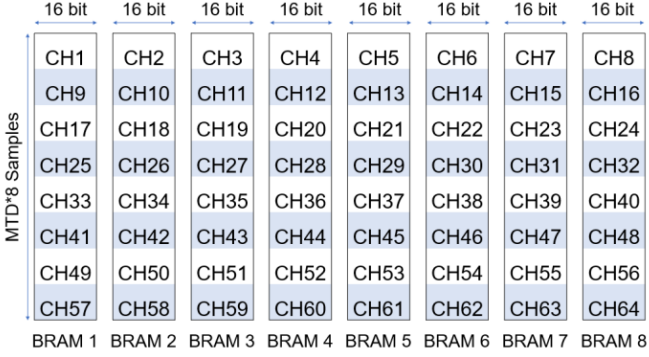

Figure 4. Diagram of the proposed transmit delay compensation implementation with BRAM. This example diagram is drawn with a setting that has $W = 64$, $F = 32$ and $F_{sub} = 4$. As a result, only 8 BRAMs each with a depth of MTD*8 instead of 64 BRAMs each with a depth of MTD are needed to implement the transmit delay compensation.

Then, the transmit delay needed to be compensated for each receive channel according to Eq. 6. As the transmit delay was different across all the receive channels, the direct way of implementing the transmit delay compensation was to use $W$ simple-two-port (S2P) block RAM (BRAM) which has one port for read and one port for write. Each BRAM buffers one receive-channel by setting the depth of each BRAM to the *maximum transmit delay* ($MTD$) where $MTD$ is the maximum of $N_n^{remove}$ through all the receive channels and steering angles. For each receive channel $n$, the interpolated RF data were written to the buffer only when $N$ was larger than $N_n^{remove}$, where $N$ was the current RF sample index which is identical to all of the input channels. When $N$ was larger than $MTD$ the interpolated RF data were read from the all the buffers simultaneously. In this way, the beginning $N_n^{remove}$ samples were removed from each channel of the interpolated RF data. Fig.3 demonstrates this process.

As the beamforming process of one row will be separated into multiple clock cycles, which will be explained in detail later, only $\frac{W*F_{sub}}{F}$ RF samples are needed by the next stage during each clock cycle, where $F_{sub}$ is the number of RF data buffers used. We can reshape the buffering of receive channel from $MTD$ by $W$ to $MTD*\frac{F}{F_{sub}}$ by $\frac{W*F_{sub}}{F}$. For each buffer, $\frac{F}{F_{sub}}$ pointers were used to track the writing address of each receive channel. In this way, each BRAM handled the transmit delay compensation for $\frac{F}{F_{sub}}$ receive channels and only $\frac{W*F_{sub}}{F}$ BRAM were needed instead of $W$ BRAM. Basically, we implemented $\frac{F}{F_{sub}}$ time interleaved first input first output (FIFO) on each of the BRAMs. The total memory space utilized by the transmit delay compensation was still $MTD*w*16\ bits$, but the number of BRAM used was reduced by a factor of $\frac{F}{F_{sub}}$.

Though each BRAM was $MTD*\frac{F}{F_{sub}}$ in depth instead of $MTD$, the Xilinx BRAM can be 1024 or 2048 samples in depth when the width was 16 bits and, in most situations, $MTD*\frac{F}{F_{sub}}$ was smaller than the maximum depth of the BRAM. Hence, we reduced the total amount of BRAM consumption by improving the utilization efficiency. An example implementation ($W = 64, F = 32, F_{sub} = 4$) is shown in Fig. 4.

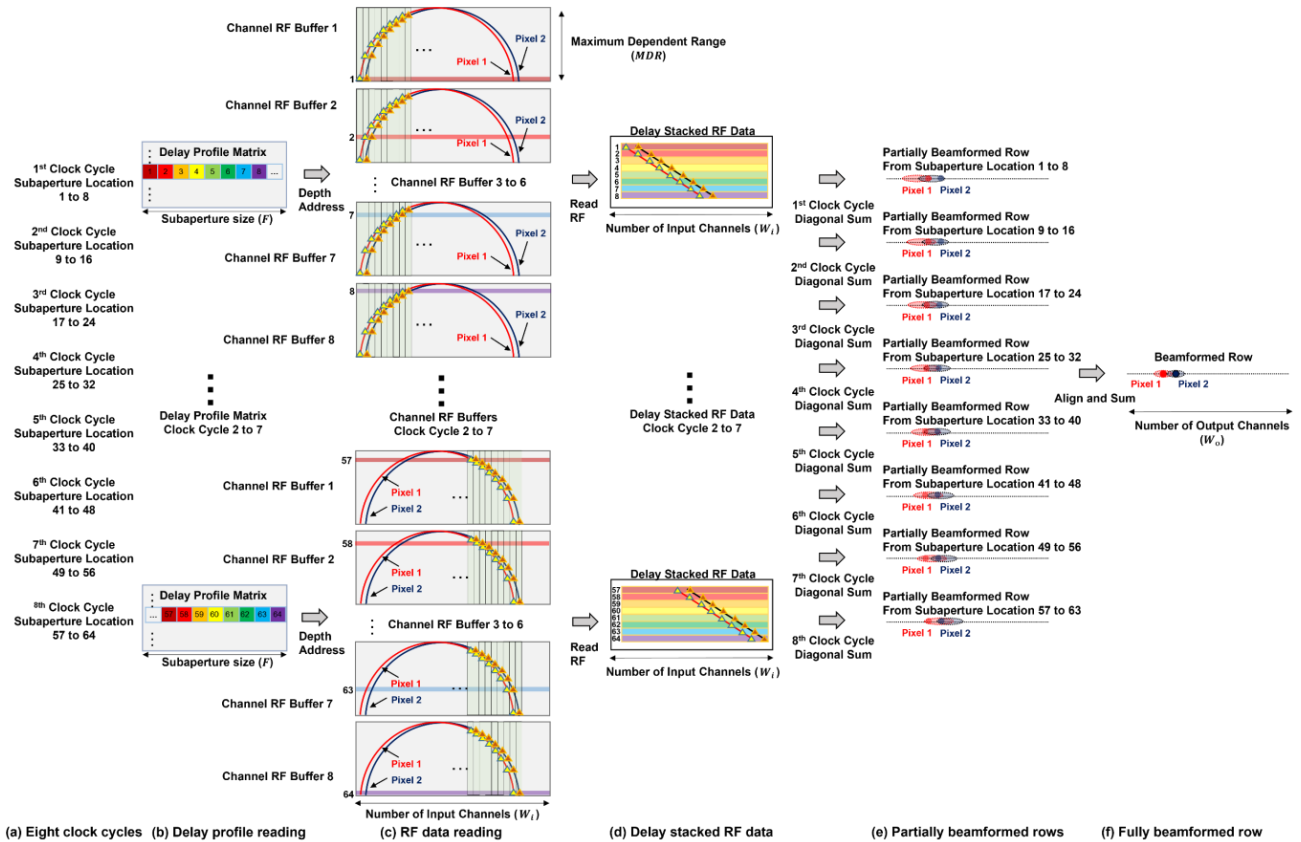

Figure 5. Diagram of the proposed beamformer design with $F_{sub} = 8$. (a) The beamforming of one row is divided into eight clock cycles. (b) Eight delay indexes are read from the delay profile matrix each clock cycle. (c) Eight rows of RF data are read from eight RF data buffers individually. (d) Delay stacked RF data matrices are formed with eight rows of RF data. (e) Partially beamformed rows are formed by diagonally summing of the delay stacked RF data matrices. (f) Beamformed row is formed by aligning and summing eight partially beamformed rows.

The receive delay compensation is a more challenging task than transmit delay compensation because the transmit delay is constant for each receive channel given a transmit angle $\theta$, while the receive delay is different at different depths. The major limitation on the hardware implementation of the beamformer to programmable logic is the total internal memory size that is needed to buffer the RF data for receive delay compensation. As discussed in the previous section we need to simultaneously read multiple rows of RF data from the RF data buffers which are implemented by the FPGA's internal memory (e.g., BRAM). If we assume each frame of RF data contains 128 channels ($W = 128$), and each channel has 2,560 samples in depth ($D = 2560$), then each RF buffer's data size is 5 Mb if the quantization depth is 16 bits (e.g., 128 x 2,560 x 16 bits = 5 Mb). If the subaperture size is 64 ($F = 64$), and we want to finish the beamforming in one clock cycle, then we need 64 RF buffers to read 64 rows of RF data simultaneously assuming each RF buffer provides a single read port (only one row could be read every clock cycle). The total memory size could reach 320 Mb (e.g., 5 Mb x 64 = 320 Mb), which is much larger than the capacity of most current FPGAs.

To address this challenge and make the implementation practical, two strategies were utilized. First, instead of buffering the whole frame of RF data to the internal buffer, only a portion of the RF data was buffered, because each row of the beamformed sample only needed RF data over a limited range of depth, which was defined as dependent range (DR), and the largest DR in all the rows of one frame was defined as maximum dependent range (MDR). Obviously, MDR was much smaller than the total depth of one frame.

If the subaperture size ($F$) is fixed, the DR is larger for shallower regions than that for deeper regions where the delay profile is flatter, which is described by the derivative of $\tau_{total}(\theta, \Delta x, z)$ with respect to $\Delta x$,

$$\frac{\partial \tau_{total}(\theta, \Delta x, z)}{\partial \Delta x} = \frac{\Delta x}{c * \sqrt{\Delta x^2 + z^2}} - \frac{\sin \theta}{c}. \qquad (9)$$

To further reduce MDR and save internal memory resources, the subaperture size was reduced for the shallower regions (close to the probe surface) in a fixed F-number way to maintain a homogenous lateral resolution.

As only a portion of the RF data were needed for the beamforming of each row of image, the size of the RF buffer could be reduced from the total depth to MDR by using cyclic buffers. To be specific, a modulo operator with divisor equal to MDR was added to the address port of RF data buffers.

As a result, the buffer depth, which is the same as MDR, was reduced from 2,560 samples to 150 samples with a Verasonics L35-16vX probe and a fixed F-number of unity. Under this condition, the size of each RF buffer was reduced to 300 Kb (128 x 150 x 16 bits = 300 Kb).

Second, instead of using subaperture size number of RF buffers, we used fewer RF buffers ($F_{sub} \le F$) and separated the beamforming of one row to multiple clock cycles, which is equal to $\frac{F}{F_{sub}}$. In each clock cycle, only $F_{sub}$ channels of the subaperture were beamformed. Fig. 5(a) shows the case of $F_{sub} = 8$. Eight RF data depth indices are read during each clock cycle from the delay profile matrix, as shown in Fig. 5(b). Only eight rows of RF data need to be read each clock cycle. As RF data buffers were configured as one port for read and one port for write, eight RF buffers were needed, as shown in Fig. 5(c). The total RF buffer size was reduced to 2,400 Kb (8 x 300 Kb = 2400 Kb), which was less than 1% of storage consumption for the unoptimized case and was an acceptable value for practical FPGA implementation. Then, the delay stacked RF data matrix was formed based on eight rows of RF data, as shown in Fig. 5(d). The partial beamformed row from eight rows of RF data is shown in Fig. 5(e). Next, the partial beamformed rows from eight clock cycles were aligned and summed up to form the beamformed row, as shown in Fig. 5(f). It generates the same results as the section IIC, i.e., one row is beamformed in one clock cycle, but with fewer resources demanded by increasing the overall processing time.

As a result, we can perform the beamforming of one row in only eight clock cycles as we fix the subaperture size at 64 elements. The beamforming time of one frame is 2,560 * 8 = 20,480 clock cycles. The RF data input and beamformed data output occur simultaneous to the beamforming process because we pipeline the whole process.

### E. Scalability

As the previous section explained, the implementation of the proposed beamformer is parameterized by several key variables, such as $W$, $F$ and $F_{sub}$. In the previous section, we assume the number of output A-lines $W_o$ is the same as the number of elements $W_i$ and use $W$ to represent both parameters to simplify the discussion. To have a finer lateral grid on the output beamformed image, multiple identical beamformers with different delay profiles that are used could be implemented. The number of identical beamformers that are implemented is defined as $R$. With the introduction of $R$, the number of output A-lines $W_o$ is redefined as $W_i * R$.

By tuning these four parameters ($W_i$, $F$, $F_{sub}$ and $R$), we can generate different FPGA implementations to adapt to different beamforming speeds and FPGA resource requirements by a single click inside the Xilinx HLS without manually rewriting the code. In this paper, we explored four different settings, which are listed in Table I, to adapt to different FPGA models and beamforming parameters. Setting 1 is appropriate for handheld ultrasound with strict power limitations. Setting 2 is appropriate for general ultrafast ultrasound imaging. Settings 3 and 4 are appropriate for high-end ultrafast ultrasound imaging with either extreme frame rate or higher lateral resolution needs.

**Table I. Design scaling settings**

| Setting | Device | $W_i$ | $F$ | $F_{sub}$ | $R$ |
|---------|--------|-------|-----|-----------|-----|
| 1 | ZU5EV | 64 | 32 | 4 | 1 |
| 2 | ZU9EG | 128 | 64 | 8 | 1 |
| 3 | KU19P | 128 | 64 | 16 | 1 |
| 4 | KU19P | 128 | 64 | 8 | 2 |

### F. Interface

In the proposed design, the beamformer was packaged as an Intellectual Property (IP) core which was automatically generated by the Xilinx Vitis HLS to be easily migrated. The AXI4 interface is a memory mapped interface, which can be connected to the DDR memory to read the raw RF data from the DDR memory and write the beamformed results back to the DDR memory. Each AXI4 port could provide up to 156 Gbps bandwidth when it is connected to a 64-bit 2400 MHz DDR4 memory. The DDR memory can be operated as a large data buffer to amend the gap between the peak data rate of input data, which can be as large as 112 Gbps for a 128 channel system with 14-bit, 62.5 MHz ADC and the sustainable average beamforming rate of the beamformer. After loading the data from the external DDR through the AXI4 interface, the data were loaded to the internal BRAMs by the beamformer.

## III. EXPERIMENT SETUP

### A. Design realization

The proposed method was written by C++ and synthesized to Verilog by Xilinx Vitis High-Level Synthesis (HLS 2022.2). Xilinx pragmas were used to instruct HLS to apply the parallelization to the C++ code. This workflow enabled fast implementation and verification of the proposed FPGA design. The synthesized results were then implemented by Xilinx Vivado ML edition 2022.2.

### B. Data sets

We used the raw RF channel data recorded from different arrays connected to a Verasonics Vantage system as data input to the beamformer. The beamformed IQ data from the Verasonics and the resulting images were compared to the results obtained using our beamformer.

A tissue-mimicking phantom (CIRS 040GSE) was scanned with a Verasonics L11-5v probe to evaluate lateral resolution and contrast. In this set, the imaging was performed using plane-wave compounding with steering angles from $-18°$ to $18°$ with a step size of $6°$. A transmit frequency of 7.8 MHz and an RF sampling rate at 31.25 MHz were used in this scan.

As a second test, a mouse brain was scanned with a Verasonics L35-16vX probe. The mouse was injected with microbubbles (Definity, Lantheus Medical Imaging, Inc.) to conduct super-resolution ULM of the mouse brain and to evaluate the speed of our beamformer. Approval of all ethical and experimental procedures was granted by the Institutional Animal Care and Use Committee (IACUC) at the University of Illinois Urbana-Champaign (Protocol No. 19063). The mouse was anesthetized by 4% isoflurane mixed with the medical oxygen in a gas induction chamber. The mouse was then transferred to the customized imaging stage and the mouse head was fixed to the stereotaxic frame with ear bars. Furthermore, the anesthesia was maintained by supplying the 2% isoflurane with oxygen through a nose cone. The scalp was removed and both side of the skull between Bregma and Lambda was opened using a rotary Dremel tool to expose the brain. The ultrasound transducer was placed above the cranial window with a coupling gel to image in the coronal plane. A 30-gauge catheter was cannulated through the tail vein. Then, the microbubbles, $6 \times 10^8$, were continuously infused using a programmable syringe pump (New Era Pump Systems Model 1000) at a flow rate of 10 μL/min. Ultrasound imaging was performed using nine steering angles ($-4°$ to $4°$ in $1°$ step size) with a post-compounding frame rate of 1,000 Hz. A total of 39 sets of 1,600 frames were acquired. Data in each acquisition had 1,280 samples in the axial dimension. A transmit frequency at 20 MHz and an RF sampling rate at 125 MHz were used in this scan.

The RF data and pre-calculated delay profile were then imported to Xilinx HLS C/RTL co-simulation to get the beamformed results.

## IV. RESULTS

### A. Latency and resources utilization

The latency of beamforming one frame with 1,280 samples (pre-interpolation raw RF samples) in fast imaging direction without delay profile loading time is listed in Table II. The maximum operating frequency based on the implementation results is also listed in Table II, along with the corresponding frame rate at this frequency. The number of clock cycles used to beamform one image with $F_{sub} = 8$ in setting 3, which was the same as the example situation in Section II.C, was 20,503 clock cycles. The extra 23 clock cycles were from the pipeline delay. The delay profile only needed to be loaded once before the beamforming process, as we kept the same parameters for the whole process. Therefore, the delay profile loading time would not slow down the beamformer frame rate.

**Table II. Latency and frame rate**

| Setting | Device | Latency | Clock Frequency | Frame Rate | Input Rate |
|---------|--------|---------|-----------------|------------|------------|
| 1 | ZU5EV | 20,500 | 314.26 MHz | 15,330 | 1.26 GSPS |
| 2 | ZU9EG | 20,503 | 304.79 MHz | 14,865 | 2.44 GSPS |
| 3 | KU19P | 10,259 | 302.30 MHz | 29,466 | 4.83 GSPS |
| 4 | KU19P | 20,501 | 302.76 MHz | 14,768 | 2.42 GSPS |

According to Table II, the highest sustainable average input rate among the four settings is 4.83 GSPS ($W_i * D * FPS = 128 * 1280 * 29466 = 4.83 \ GSPS$), which needs the DDR memory to provide an external memory bandwidth of 77.28 Gbps (4.83 GSPS * 16 bits). A 64-bit 2400MHz DDR4 memory could provide 153.6 Gbps (2400 MHz * 64 bits) memory bandwidth, which is enough to support the beamformer input rate.

The post-implementation hardware resource utilization and power consumption estimation from Vivado is listed in Table III.

**Table III. Resource utilization and power consumption**

| Setting | Device | LUT | BRAM | Total on-chip power |
|---------|--------|-------|------|---------------------|
| **1** | ZU5EV | 16,043 | 67 | 2.066 W |
| **2** | ZU9EG | 41,428 | 318 | 5.385 W |
| **3** | KU19P | 57,340 | 491 | 12.968 W |
| **4** | KU19P | 73,843 | 478 | 12.173 W |

According to our measurement, the Verasonics beamformer (Intel Xeon W-2155 10 cores 20 threads 3.3 GHz 64 GB RAM) achieved a frame rate of 406 fps on the same data set as setting 2 used. The proposed FPGA beamformer could speed up the beamforming by a factor of 36.6 under setting 2 and 72.6 under setting 3.

### B. Beamformed images comparison

The beamformed images from the CIRS phantom using the proposed FPGA beamformer are shown in Figs. 6 and 7. The Verasonics beamformed image of wire targets is shown in Fig. 6(a) and the FPGA beamformed image is shown in Fig. 6(b). Visually, no lateral resolution degradation of the FPGA beamformed image compared to Verasonics beamformed image was observed. The lateral resolution comparison of Verasonics and FPGA beamformer results is shown in Fig. 6(c).

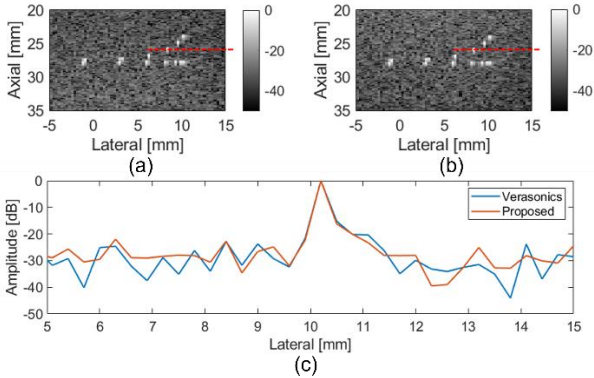

Figure 6. B-mode image of CIRS phantom wire target region with Verasonics (a) and FPGA beamformer (b). ROIs (Region of Interests) are marked with red lines. Lateral resolution comparison between Verasonics beamformer and proposed FPGA beamformer with CIRS phantom (c).

Images of an anechoic target inside the CIRS phantom were constructed using the Verasonics and the FPGA beamformer. Visually, no differences were observed between the Verasonics beamformed images and FPGA beamformed images in Fig.7. The contrast to noise ratio (CNR) was calculated to compare the performance. The CNR is given by

$$CNR = \frac{|u_i - u_o|}{\sqrt{\sigma_i^2 + \sigma_o^2}}, \quad (10)$$

where $u_i$ and $u_o$ represent the mean pixel intensity inside and outside the anechoic cyst; and $\sigma_i^2$ and $\sigma_o^2$ denote the variance of pixel intensity inside and outside the anechoic cyst. The CNR values from the Verasonics beamformer and FPGA beamformer were both 1.0.

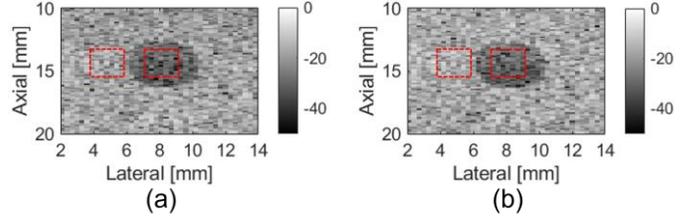

Figure 7. B-mode images of an anechoic target in the CIRS phantom with Verasonics (a) and FPGA beamformer (b). ROIs are marked with red rectangle.

Power Doppler images of a mouse brain were created by the accumulation of SVD (Singular Value Decomposition) filtered [24] 1,600 post-compounding frames. The images based on Verasonics beamformer and proposed method are shown in Fig. 8(a) and Fig. 8(b). Visually, they have comparable results in the spatial resolution of vessels. The only noticeable difference is the top part of the image results from FPGA beamformer is slightly darker than that from the Verasonics beamformer. The reason behind this is the subaperture size is smaller for the top part of image results from the FPGA beamformer, which is described in part IID.

ULM images of a mouse brain were generated by 39 sets of data described in IIIB. In the ULM processing, the MB signal with different speed ranges and directions were separated into three groups using 3D Fourier domain filters and processed separately [25]. Velocity maps were generated using a bipartite graph-based MB pairing and tracking algorithm [26],[27],[28],[29]. The final localization and velocity images were the combination of the individual reconstruction images generated from each acquisition. The resulting ULM directional flow images using both the Verasonics and FPGA beamformer are shown in Figs. 8(c) and 8(d), respectively. The zoomed in version images are shown in Figs. 8(f) and 8(g). The cross-section comparison between the directional flow images with Verasonics and FPGA beamformer is shown in Fig. 8(e).

No visually apparent difference can be identified from the ULM images constructed from Verasonics beamformer and FPGA beamformer. The cross-section comparison between the two beamformers also shows the similarities between the results of two beamformers.

## V. DISCUSSION

A new beamforming parallelization method was proposed that integrated with hardware (programmable logic implementation on an FPGA) to achieve ultrafast beamforming for ultrafast ultrasound imaging. The proposed method solves the major obstacle of achieving higher beamforming frame rate by enabling delay profile reuse and parallel beamforming. HLS based design flow also enables fast adaption to different applications and platforms. A sustainable average beamforming rate of 4.83 GSPS in terms of input raw RF samples or 29,466 FPS in term of frame rate was achieved by the proposed ultrasound PWI beamformer.

The image quality of the proposed FPGA beamformer was similar to that of Verasonics beamformer as quantified by the CNR and lateral resolution in the phantom experiments. This indicates that the proposed FPGA beamformer did not sacrifice the quality of the image for speed. The performance of the FPGA beamformer with *in vivo* data was also assessed by imaging a mouse brain injected with microbubbles. Power Doppler images and ULM images created with the FPGA beamformer were compared with images created using the Verasonics beamformer side-by-side. Cross-section plots of the ULM images were also provided to directly compare the performance. The correctness and quality of the proposed beamformer was verified by the visual similarities between the results from the proposed method and the Verasonics beamformer.

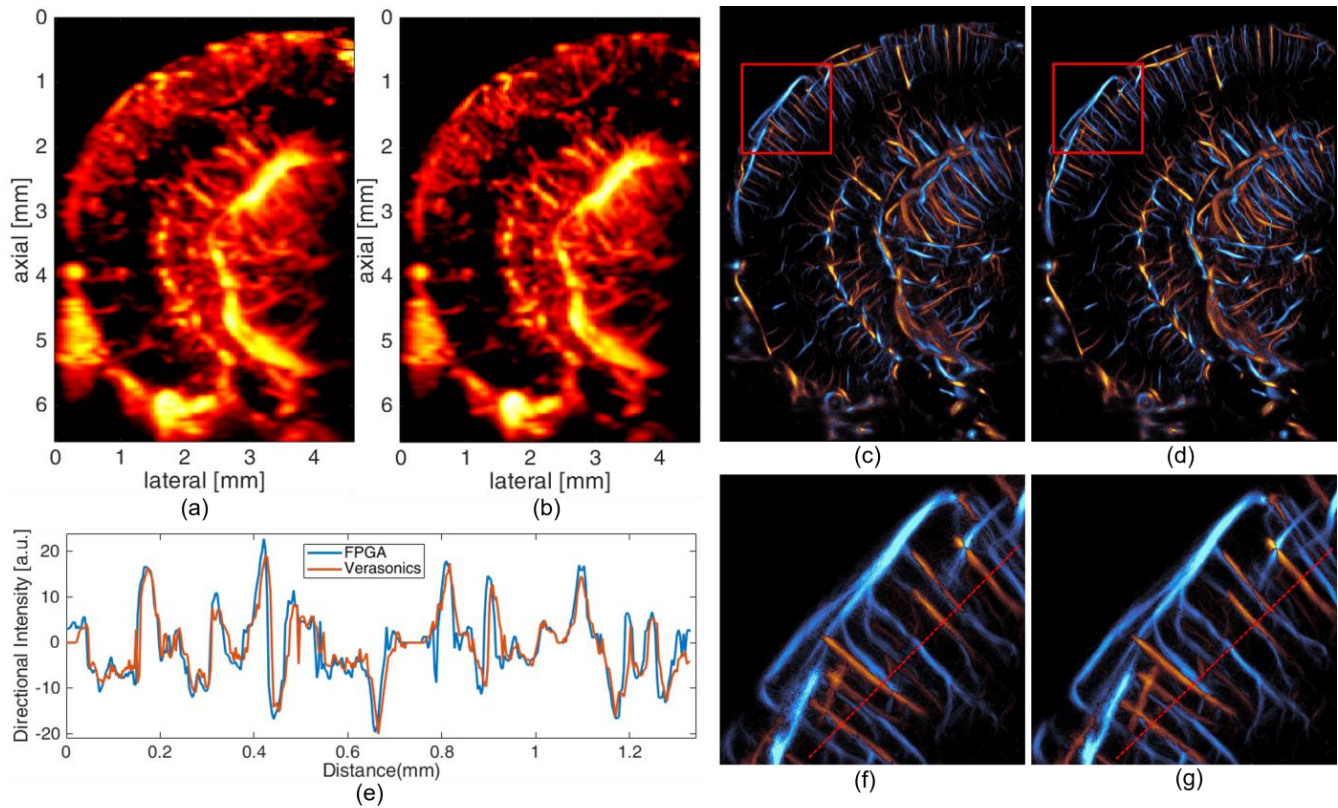

**Figure 8. Power doppler image of mouse brain with Verasonics beamformer (a) and FPGA beamformer (b). Directional flow image of mouse brain with Verasonics beamformer (c) and FPGA beamformer (d). Zoomed in version of directional flow image (marked with red rectangle) with Verasonics beamformer (f) and FPGA beamformer (g). The cross-section comparison of directional flow image (e) with Verasonics and FPGA beamformer.**

**Table IV Comparison with previous work. Schemes 1-4 represent the proposed beamformer and the last two rows are comparison beamformer schemes.**

| Scheme | Subaperture Size | RF Frame Size | Frame Rate | LUT | BRAM (blocks) | DSP (slices) |
|--------|------------------|---------------|------------|-----|---------------|--------------|
| **1** | 32 | 1,280*64 | 15,330 | 16,043 | 67 | 0 |
| **2** | 64 | 1,280*128 | 14,865 | 41,428 | 318 | 0 |
| **3** | 64 | 1,280*128 | 29,466 | 57,340 | 491 | 0 |
| **4** | 64 | 1,280*128 | 14,768 | 73,843 | 478 | 0 |
| **[17]** | 32 | 1,280*96 | 3,800 | 144,800*4 | 1,087*4 | 322*4 |
| **[18]** | 64 | 1,024*64 | 14,000 | 67,375*8 | 242*8 | 492*8 |

The importance of the proposed method consists of potentially enabling continuous unblocked ultrafast ultrasound imaging. This is possible because the proposed beamformer can be easily integrated to FPGAs that can be directly connected to AFEs without the need of PCI-Express interface and host computer in between. The bandwidth of data being transferred to a host PC could be reduced to a much lower value that could match the speed for saving to a hard drive by adding compounding and IQ demodulation to the beamformer. In this way, long duration, continuous ultrafast ultrasound can be achieved with the proposed beamformer. With continuous unblocked ultrafast ultrasound, ULM could be improved by having a larger number of frames in one data set and much longer tracking duration compared to current ultrasound research platforms. fUS could also be improved by having continuous real-time ultrafast ultrasound imaging. Furthermore, other time-consuming workload computations traditionally done on a CPU or GPU, such as compounding, IQ demodulation and high pass filtering, could be moved to an FPGA to enable faster frame rates and real-time processing.

Using the Xilinx HLS toolchain, the proposed beamformer design can be easily scaled up or scaled down to fit different platforms and applications in a short time by simply modifying several parameters.

The proposed parallelization method improved the memory utilization efficiency and simplified the memory architecture by delay profile reuse. All the summing operations in the proposed design were performed with LUT instead of DSP cores to save DSP resources for other operations that needed multiply operations and to lower the power consumption. As a result, the low power consumption feature of the proposed FPGA-based beamformer makes ultrafast ultrasound imaging possible on portable devices by reducing the beamformer power consumption to about two watts. Dynamic apodization, which can efficiently reduce sidelobes, is a feature that has not been considered in the proposed method. It is a tradeoff between the resource utilization and the sidelobe level we chose. Then, the dynamic apodization needs DSPs to perform multiplications, companion LUTs and memory resources to load and buffer the apodization coefficient, and an increase in power consumption due to more resource utilization. A preliminary experimental implementation with setting 2, plus dynamic apodization, resulted in a 70% increase in the number of LUTs and 40% increase in power consumption along with utilization of over 900 DSPs.

The processes after the DAS, such as compounding and digital down conversion, were not implemented in the proposed design as these are beyond the scope of beamforming. In the future complete ultrafast ultrasound scanning system, these two functions will be implemented by accumulating the beamramed results on a frame buffer using internal Xilinx Ultra RAM (URAM) which has larger size than BRAM. Then, the digital down convertor can be generated by traditional DSP design tools such as Xilinx System Generator and imported to HLS using the generated RTL code.

A comparison of this work with previous FPGA-based beamformers is shown in Table IV. The resource utilizations of previous works have been converted to the same standard for easy comparison. One thing to notice is that the previous works are all complete real-time ultrasound scanning systems, which includes more processing and control in the FPGA that consumes more FPGA resources than the proposed design, which is currently limited to simulation. As a result, the resource utilization in Table IV only serves as a reference instead of a fair comparison.

Also, the proposed methods were designed specifically for ultrafast plane wave imaging and are not applicable to focused imaging or divergent wave imaging which may limits its range of usage. The tradeoff was selected to achieve higher beamforming rates within a compact design, while the previous works are also applicable to divergent or focus wave imaging. However, in terms of performance or speed, the proposed design provides the highest beamforming rate because of delay profile reuse, which eliminates the run-time delay profile calculation and simplifies the memory reading architecture.

The proposed design has not been integrated to an ultrasound scanning system for real-time online demonstration. The real-time online implementation needs to be done with a customized ultrasound scanning system that connects all the AFEs to a single FPGA. This is feasible with newest AFE and FPGA technologies that use JESD204B interface to replace traditional LVDS interface which can greatly ease the pin counts limitation and printed circuit board layout challenges. Though the physical interface part has not been included in the simulation, a proof of concept prototype has been built to demonstrate that 16 channels of ultrasound data can be transferred to a single FPGA via only two lanes of JESD204B interfaces [30]. Xilinx JESD204B IP was used in this prototype to convert the high-speed serial interface from the AFE to a parallel interface. Current Xilinx FPGA such as Virtex Ultrascale+ provides the capabilities of connecting up to 128 lanes of JESD204B interfaces to a single FPGA, which means up to 1024 ultrasound channels can be connected to a single FPGA. This is far above the channel counts (128 or 256) of common linear array used by ultrafast ultrasound imaging. For a channel count higher than 128, multiple DDR4 memory or high bandwidth memory (HBM) should be used to provide an adequate memory bandwidth for both the ADC data and the beamformed data.

## VI. CONCLUSION

The proposed FPGA implementation of an ultrafast beamformer enabled steered PWI with high versatility and scalability. Due to the versatility of the implementation, there is no need to regenerate the bitstream or reprogram the FPGA to adapt to different probes or steering angles. The HLS allows changing the scale of FPGA beamformer to fit different application scenarios in a very short turnaround. This architecture can be utilized in both high-end ultrasound research platforms that need a frame rate of over 29,000 FPS or for portable pocket ultrasound scanners that need a high efficiency, low power compact FPGA beamformer.

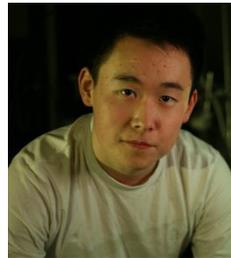

Zhengchang Kou (Graduate Student Member, IEEE) received his B.S. and M.S. degrees in electrical engineering from the University of Illinois Urbana-Champaign, Urbana, IL USA, in 2017 and 2020. He is now pursuing the Ph.D. degree in the Department of Electrical and Computer Engineering at the University of Illinois Urbana-Champaign, Urbana, IL, USA. His current research interests are ultrafast ultrasound imaging, ultrasound communication and hardware acceleration on FPGA.

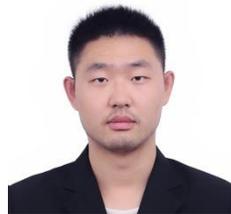

Qi You (Graduate Student Member, IEEE) received the B.S. and M.S. degree in Electronic Science and Engineering from Nanjing University, Nanjing, Jiangsu, China. He is now pursuing the Ph.D. degree in the Department of Bioengineering at the University of Illinois at Urbana-Champaign, Urbana, IL. His current research interests are ultrasound beamforming and super-resolution techniques.

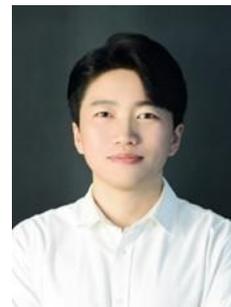

Jihun Kim (Member, IEEE) received the Ph.D. degree in the Department of Information & Communication Engineering from Daegu Gyeongbuk Institute of Science & Technology (DGIST), Daegu, S. Korea in 2019. From 2019 to 2020, he was a Post-doctoral Research Associate with the Department of Aerospace and Mechanical Engineering (A&M), University of Notre Dame, Notre Dame, IN, USA. From 2021 to 2023, he was a Post-doctoral Research Associate with the Department of Electrical and Computer Engineering (ECE), University of Illinois at Urbana–Champaign (UIUC), Urbana, IL, USA. He is currently an Assistant Professor in the Department of ICT Convergence Engineering/Electronic Engineering at the Kangnam University, S. Korea. His research interests focus on the development of deep learning-based novel super-resolution ultrasound imaging techniques.

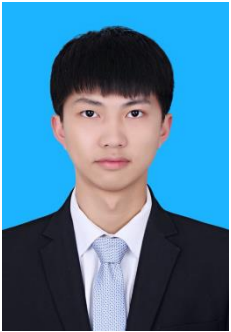

Zhijie Dong (Graduate Student Member, IEEE) received his B.Eng. degree in Information Engineering from the Southeast University, Nanjing, China, in 2017; and the M.S. degree in Electrical and Computer Engineering from the University of Michigan, Ann Arbor, MI, in 2018. He is currently pursuing the Ph.D. degree in the Department of Electrical and Computer Engineering at the University of Illinois at Urbana-Champaign, Urbana, IL. His current research interests are ultrafast 3D ultrasound imaging and deep learning in ultrasound.

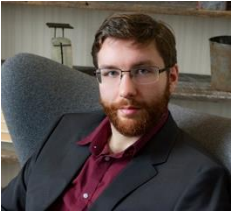

Matthew Lowerison received the Ph.D. degree in Medical Biophysics from the University of Western Ontario, London, Ontario, Canada in 2017. He is now with the Department of Electrical and Computer Engineering and the Beckman Institute for Advanced Science and Technology at the University of Illinois at Urbana-Champaign, Urbana, IL. His current research interests include ultrasound microvessel imaging, super-resolution ultrasound localization microscopy, and ultrasonic characterization of tumor microenvironments.

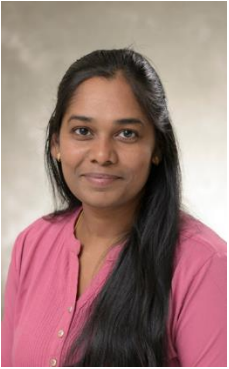

Nathiya Vaithiyalingam Chandrasekaran earned her B.S in 2006, M.S - Medical Biochemistry and Master of Philosophy in 2009 from University of Madras, India. She earned Ph.D. in Biochemistry learning Stress induced Depression from the University of Madras, India, in 2015. She works at Molecular and Integrative physiology and Beckman Institute with Professor Daniel Llano on studying aging-related auditory network function, particularly relates to its changes in network properties in the Auditory cortex and thalamus. she also study the cerebro microvasculature using Super-resolution Ultrasound Microscopy to Image the Aging Brain and Alzheimer's mouse model to understand the deeper vascular structures.

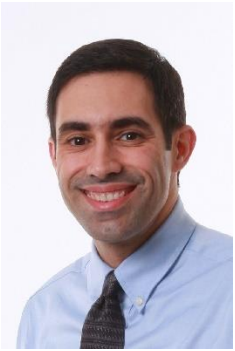

Daniel Llano obtained his MD and PhD degrees from the University of Illinois. He is currently an associate professor of Molecular and Integrative Physiology at the University of Illinois. His research is focused on systems neuroscience.

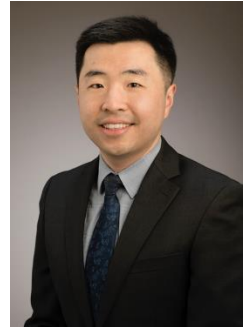

Pengfei Song (IEEE/S'09/M'14/SM'19) received the Ph.D. degree in Biomedical Engineering from Mayo Clinic College of Medicine in 2014, under the supervision of Drs. James Greenleaf and Shigao Chen. He is currently an Assistant Professor in the Department of Electrical and Computer Engineering, and the Beckman Institute for Advanced Science and Technology at the University of Illinois Urbana-Champaign. His current research interests are super-resolution ultrasound imaging, ultrafast 3D imaging, deep learning, functional ultrasound, and ultrasound shear wave elastography. Dr. Song is the recipient of the NIH K99/R00 Pathway to Independence Award, the NSF CAREER Award, and the NIH/NIBIB Trailblazer Award. He has authored and co-authored over 80 peer-reviewed journal articles in the field of ultrasound imaging. He holds several patents that have been licensed and commercialized by major ultrasound companies and used worldwide in the clinic. Dr. Song is a senior member of IEEE, a Fellow Member of the American Institute of Ultrasound in Medicine, a senior member of the National Academy of Inventors, and a full member of ASA.

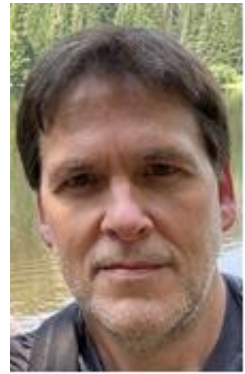

Michael L. Oelze (M'03–SM'09) was born in Hamilton, New Zealand, in 1971. He received the B.S. degree in physics and mathematics from Harding University, Searcy, AR, USA, in 1994, and the Ph.D. degree in physics from the University of Mississippi, Oxford, MS, USA. From 2000 to 2002, he was a Post-Doctoral Researcher with the Bioacoustics Research Laboratory, Department of Electrical and Computer Engineering (ECE), University of Illinois at Urbana–Champaign (UIUC), Urbana, IL, USA. From 2002 to 2004, he was an NIH Fellow conducting research in quantitative ultrasound techniques for biomedical ultrasound applications in cancer detection. In 2005, he joined the Faculty of ECE, UIUC. He is currently a Professor and the Associate Head for Graduate Affairs in ECE, a Health Innovator Professor in the Carle Illinois College of Medicine and the Frederick G. and Elizabeth H. Nearing Scholar, Grainger College of Engineering. His research interests include biomedical ultrasound, quantitative ultrasound imaging for improving cancer diagnostics and monitoring therapy response, ultrasound tomography, ultrasound-based therapy, beamforming, coded excitation, communications using ultrasound, in body devices, and hardware solutions. Dr. Oelze is currently a Fellow of the AIUM, and a member of ASA. He is a member of the Technical Program Committee of the IEEE Ultrasonics Symposium. He currently serves as an Associate Editor for IEEE TRANSACTIONS ON ULTRASONICS, FERROELECTRICS, AND FREQUENCY CONTROL, Ultrasonic Imaging, and IEEE TRANSACTIONS ON BIOMEDICAL ENGINEERING.